\documentstyle[12pt]{article}
\begin{document}
\title{ Scattering of Glueballs and Mesons in Compact $QED$ in $2+1$
Dimensions }
\author{ A.M. Chaara$^{1}$, H. Kr\"oger$^{1,2}$,
L. Marleau$^{1}$, K.J.M. Moriarty$^{3}$ and J. Potvin$^{4}$ \\ [2mm]
{\small\sl $^{1}$ D\'epartement de Physique, Universit\'e Laval,
Qu\'ebec, Qu\'{e}bec G1K 7P4, Canada } \\
{\small\sl $^{2}$ Max Planck Institut f\"ur Kernphysik,
P.O.Box 10 39 80, D-6900 Heidelberg, Germany } \\
{\small\sl $^{3}$ Department of Mathematics,
Statistics and Computing Science,
Dalhousie University,} \\
{\small\sl Halifax, Nova Scotia B3H 3J5, Canada } \\
{\small\sl $^{4}$ Department of Science and Mathematics,
Parks College of St. Louis University,} \\
{\small\sl Cahokia, Illinois 62206 USA }}
\date{ LAVAL-PHY-94-05 \\
PARKS-PHY-94-03 }
\maketitle
\begin{flushleft}
{\bf Abstract}
\end{flushleft}
We study glueball and meson scattering in compact $QED_{2+1}$ gauge theory
in a Hamiltonian formulation and on a momentum lattice.
We compute ground state energy and mass, and
introduce a compact lattice momentum operator
for the computation of dispersion relations.
Using a non-perturbative time-dependent method we compute
scattering cross sections for glueballs and mesons.
We compare our results with strong coupling perturbation theory.
\begin{flushleft}
PACS index: 11.15.Ha
\end{flushleft}
\setcounter{page}{0}
\maketitle

\newpage
\paragraph{1. Introduction.}
The non-perturbative computation of scattering and decay
amplitudes of hadrons such as mesons and glueballs from
quantum chromodynamics ($QCD$) is one of the most challenging
problems of nuclear physics.
Recently, hadron-hadron scattering lengths have been obtained
from ${QCD}_{3+1}$ lattice simulations by studying the finite volume
behavior of the energy of a two-particle state
\cite{kn:Hamb83}.
The $\pi-\pi$ and the $N-N$ scattering length has
been obtained in this
way by Guagnelli et al. \cite{kn:Guag90}.
Sharpe et al. \cite{kn:Shar92}
have computed the $\pi-\pi$ scattering length in the $I=2$ channel.
A method of how to obtain scattering phase shifts from the finite
volume behavior of the spectrum has been worked out by L\"uscher
\cite{kn:Lusc86}.
Meson-meson phase shifts from compact ${QED}_{2+1}$ have also been
obtained in lattices simulations using L\"uscher's method by Fiebig et al.
\cite{kn:Fieb92}.
All these studies have used the powerful numerical
methods of Euclidean field theory.
Scattering can also be studied using
Hamiltonian field theory, such as in the work of
Alessandrini et al.\cite{kn:Ales80} and by Hakim \cite{kn:Haki81}, who have
considered glueball scattering in strong coupling perturbation theory
to lowest order in compact ${QED}_{2+1}$.

At present it seems difficult
to extend these Hamiltonian calculations to higher orders of perturbation
theory, or to apply it to non-Abelian gauge theory.
On the other hand, L\"uscher's method applied to $QED$ features ambiguities in
selecting the physical phase shifts from a determinant.
These difficulties justify trying alternative methods to
compute directly the scattering matrix in lattice gauge theory.
Such a non-perturbative Hilbert space
method to study scattering in quantum field theory on the lattice has been
developped by Kr\"oger and collaborators (see  Ref. \cite{kn:Krog92} and
references therein). It has been
applied to the Schwinger model on the lattice \cite{kn:Brie89}.
Here we will present an application to $QED_{2+1}$
and compute scattering cross sections.

This paper will be organized as follows: First, we briefly describe
compact lattice $QED$ in 2+1 dimensions.
We present results for the low-lying
mass spectrum and the dispersion relation for
the glueball and meson and finally S-matrix elements and
cross sections for glueball-glueball
and meson-meson scattering.\\

\paragraph{2. Compact $U(1)$ gauge theory in 2+1 dimensions.}
As shown by Polyakov \cite{kn:Poly77} and Banks et al. \cite{kn:Bank77}
compact $U(1)$ gauge theory in 2+1 dimensions has many properties
of $SU(3)$ gauge theory in 3+1 dimensions, such as field-source confinement
for every non-vanishing value of the coupling constant.
Ambjorn et al. \cite{kn:Ambj82} found that the mass gap and the string tension
vanish at $g=0$ in a non-analytic way, that is,
\begin{eqnarray}
&  & M^{2} = \frac{8 \pi^{2}}{g^{2}a^{3}} \exp \left[ \frac{ -2 \pi^{2} V(0) }
{g^{2} a} \right],
\nonumber \\
&  & \sigma = \frac{2}{\pi^{2}} g^{2} M.
\end{eqnarray}
G\"{o}pfert and Mack \cite{kn:Gopf82} have shown rigorously that the
theory has a non-zero string tension for arbitrary $g \neq 0$ and that
the value of $\sigma$ given by eq.(1) is a rigorous lower bound on the
string tension.
An exact ground state solution for this model in a Hamiltonian lattice
formulation has been obtained by Shuohong et al. \cite{kn:Shuo88}. Other
quantities such as the mass gap and the string tension have been obtained
at high orders in weak coupling perturbation theory by Morningstar
\cite{kn:Morn92}.
This model has been investigated using weak coupling and strong coupling
perturbation theory, variational methods, and Monte Carlo methods,
to compute the vacuum energy, glueball masses, string tension,
specific heat, etc.

We start with the usual Kogut-Susskind
Hamiltonian for the compact $U(1)$ model in 2+1
dimensions, using the temporal gauge,
with rescaled variables following Drell et al.
\cite{kn:Drel76}. To compare lattice results with
strong coupling perturbation theory,
we rescale $\frac{2}{g^{2}} H \rightarrow H$
\cite{kn:Irvi83} and work with the following Hamiltonian,
\begin{equation}
H =  H_{0} + H_{int} =
\sum_{links} (E_{\vec{n}}^{x})^{2} + (E_{\vec{n}}^{y})^{2}
- x \sum_{plaq} V_{\vec{n}} + V_{\vec{n}}^{\dag},
\end{equation}
where $x = 1/g^{4}$ is the dimensionless coupling parameter.
Starting from the vector potential $A^{j}_{\vec{n}}$ with $j=x,y$
and $\vec{n}$ denoting the lattice site, then a link variable is
defined by $U_{\vec{n}}^{j}=\exp [iA_{\vec{n}}^{j}]$, and a plaquette variable
being the product of four sequential links going around an
elementary square tile is defined
by $ V_{\vec{n}} = U^{1}_{\vec{n}} \; U^{2}_{\vec{n}+\hat{1}} \;
[U^{1}_{\vec{n}+\hat{2}}]^{\dagger} \; [U^{2}_{\vec{n}}]^{\dagger}$.
The Hamiltonian is built from link and plaquette variables
and it is invariant under residual, i.e., time independent gauge
transformations.
The Hamiltonian shows a number of symmetries:
(a) The Hamiltonian can create or annihilate a single plaquette, but
it {\em cannot} create or annihilate a single link. Thus an invariant subspace
of Hilbert space is formed by those states which are composed of multiple
plaquettes. There are other invariant subspaces built from , e.g.,
two links and multiple plaquettes.
(b) The Hamiltonian is invariant under simultaneous change of
direction of all links and plaquettes. This allows one to distinguish between
symmetric and antisymmetric states.
(c) The Hamiltonian conserves the total lattice wave vector. For example, if
one
forms a state on the $\vec{k}$-lattice by discrete Fourier transformation,
\begin{equation}
\mid \phi(\vec{k}_{a}, \cdots , \vec{k}_{z} ) >
= \sum_{\vec{a}, \cdots , \vec{z}}
\exp \left[ i \vec{k}_{a} \cdot \vec{a} + \cdots +
i \vec{k}_{z} \cdot \vec{z} \right]
V_{\vec{a}} \cdots V_{\vec{m}} \;
V_{\vec{n}}^{\dagger} \cdots V_{\vec{z}}^{\dagger}
\mid \{ 0 \} >,
\end{equation}
then the Hamiltonian leaves invariant the subspace of those states with
$\vec{k}_{a} + \cdots + \vec{k}_{z} = \vec{k}_{tot} = \mbox{const}$. Let
us denote this subspace by ${\cal H}(\vec{k}_{tot})$.
Below we will indicate the relationship between the total
wave vector $\vec{k}_{tot}$ and the lattice momentum operator $\vec{P}$.
We will make use of all of these symmetries when constructing glueball-like
and meson-like states, and in particular scattering amplitudes, which conserve
the total wave vector.\\

\paragraph{3. Hilbert space and physical states.}
We choose a basis of link states which diagonalizes the electrical field,
$E_{\vec{n}}^{i} \mid \ell_{\vec{n}}^{i} > =
\ell_{\vec{n}}^{i} \mid \ell_{\vec{n}}^{i} >$.
The Hilbert space ${\cal H}$ is built from states written as a tensor product
of link states. The state $\mid \{ 0 \} >$
denotes an absent electric field on the lattice and is called
electric vacuum.

In order to describe a meson, it has been suggested by
Potvin and DeGrand \cite{kn:Potv84} to use classical heavy [static] quarks as
sources. A meson can be regarded as a state of
electric flux lines between a heavy quark and its antiquark,
separated by one or several units of electrical field on the lattice.
To lowest order in the strong coupling expansion, a meson is given by a
single link of length one between a quark and an antiquark.
To higher order, fluctuations of plaquettes will occur similar to those
arising from vacuum fluctuations in ordinary $QED$. A one-meson state obeys
an eigenvalue equation of the Hamiltonian and it
must satisfy Gauss' law, with charge distribution corresponding to one quark
and one antiquark.

A glueball state is a state without quark sources. It obeys
the eigenvalue equation of the Hamiltonian and Gauss' law,
with a charge distribution identically zero.
To lowest order in the strong coupling expansion, it is an
eigenstate of the unperturbed Hamiltonian $H_{0}$, it
consists of just one plaquette. To higher order, more plaquettes
will be added.

Similarly, the ground state of the theory
obeys the eigenvalue equation plus Gauss' law
with the eigenvalue $E_{vac}$ being the lowest eigenvalue of the spectrum.
One expects the vacuum to lie in the same sector of Hilbert space
as the glueball states, i.e. in the subspace ${\cal H}(\mbox{zero-link})$,
which means for the charge distribution $\rho \equiv 0$.
Moreover one expects the vacuum to have zero lattice wave vector, i.e.
to lie in the subspace ${\cal H}(\vec{k}_{tot}=0)$.
Both these properties have been checked in the numerical calculations
and have been found to be satisfied. To lowest order in the strong coupling
expansion, the vacuum is identical to the electric vacuum $\mid \{ 0 \} >$.

In order to map the Hamiltonian onto a finite dimensional matrix,
we have chosen to work in the sector of Hilbert space in which
states are composed of zero links and up to four plaquettes to represent a
glueball and a meson is represented by one link and up to four
plaquettes. This applies to the calculation of the mass spectrum.
In the scattering calculations we have also worked in the sector of
up to four plaquettes, i.e., glueball/meson being represented by zero/one link
and up to two plaquettes each. The explicit construction and
graphical representation of these state vectors is provided in Ref.
\cite{kn:Chaa92}.
In principle, the number of links and plaquettes limits the domain
of the strong coupling regime.
The actual domain of validity of our calculation will be studied
by comparing with similar results obtained
from strong coupling perturbation theory.\\

\paragraph{4. Lattice momentum operator and dispersion relations.}
As discussed above it is our aim to compute a scattering reaction
by simulation on the lattice.
As we intend to apply a Hilbert space method, we need to construct asymptotic
states,
which are characterized by energy, momentum and other quantum numbers.
Thus in analogy to the lattice Hamiltonian, given by eq.(2), we will construct
below a
lattice momentum operator and compute dispersion relations.
The compact lattice Hamiltonian has been constructed
by Kogut and Susskind \cite{kn:Kogu75}.
We proceed to construct the compact lattice momentum
operator in a similar way.

In general, the momentum operator in electrodynamics is given by
the Poynting vector,
\begin{equation}
\vec{P} = \int d^{2}x \; \vec{E}(x) \times \vec{B}(x).
\end{equation}
In two spatial dimensions, the magnetic field, or in other words
the curl of the gauge field $A$, is no
longer a pseudovector but now a pseudoscalar and so we write
$B \equiv B^{z}$.
On the lattice in non-compact form we have
\begin{equation}
\vec{P} = \sum_{x_{i}} a^{2} \left[ -E^{x}(x_{i}) B(x_{i}) \vec{e}_{y}
+ E^{y}(x_{i}) B(x_{i}) \vec{e}_{x} \right].
\end{equation}
One goes over to the compact form via
$B \rightarrow  \sin [e a^{2} B] / e a^{2} $.
Rescaling the variables following Drell \cite{kn:Drel76}, plus rescaling
$\vec{P}$ by $a \vec{P} \rightarrow \vec{P}$,
yields, if we write $\vec{P}$ as Hermitian operator, the compact lattice
momentum operator,
\begin{equation}
\vec{P} = \frac{1}{4 i} \sum_{\vec{n}}
\left[ E^{y}_{\vec{n}} \; [ V_{\vec{n}}
- V_{\vec{n}}^{\dagger} ] + [ V_{\vec{n}} - V_{\vec{n}}^{\dagger} ] \;
E^{y}_{\vec{n}} \right] \vec{e}_{x}
- \left[  E^{x}_{\vec{n}} \; [ V_{\vec{n}}
- V_{\vec{n}}^{\dagger} ] + [ V_{\vec{n}} - V_{\vec{n}}^{\dagger} ] \;
E^{x}_{\vec{n}} \right] \vec{e}_{y}.
\end{equation}
The compact Hamiltonian, given by eq.(2), is gauge invariant under the
residual gauge transformations compatible with temporal gauge fixing.
The generator of those gauge transformations is given by [Gauss' law]
$ \vec{\nabla} \cdot \vec{E} = \rho$ [see Ref. \cite{kn:Hell81}].
The electric field $E^{i}_{\vec{n}}$ and the plaquette $V_{\vec{n}}$ commute
with the generator of gauge transformations. Hence also the compact
lattice momentum operator $P^{i}$, given by eq.(6) is gauge invariant.
Apart from gauge invariance,
the compact lattice momentum operator has the same symmetries as the compact
lattice Hamiltonian: (a) it creates or annihilates plaquettes but not single
links, (b) it is invariant under exchange of orientation, (c) it conserves
the total lattice wave vector $\vec{k}_{tot}$.

It is obvious that the compact lattice Hamiltonian,
when reversing the scaling of the variables, goes over in the classical
continuum limit $ a \rightarrow 0$ to the standard expression,
\begin{equation}
H \rightarrow_{a \rightarrow 0}
\frac{1}{2} \int d^{2}x \; \vec{E}^{2}(x) + \vec{B}^{2}(x).
\end{equation}
In the same way also the compact lattice momentum operator,
when reversing the scaling of the variables, goes over in the classical
continuum limit to the Poynting vector, i.e. the standard expression,
\begin{equation}
\vec{P} \rightarrow_{a \rightarrow 0} \int d^{2}x \vec{E}(x) \times \vec{B}(x).
\end{equation}
The Hamiltonian and the momentum operator are part of the set of operators,
which form the Lie-algebra of the Poincar\'{e} group.
In the continuum case, one has $\left[ H_{cont}, \vec{P}_{cont} \right] =0$.
On a finite lattice [$ a \neq 0 $] the commutator
for the compact lattice operators yields
$\left[ H_{latt}, \vec{P}_{latt} \right] \neq 0$.
However, one has
$\left[ H_{latt}, \vec{P}_{latt} \right] \rightarrow_{a\rightarrow 0} 0$.

When trying to compute numerically dispersion relations of
the compact lattice Hamiltonian $<H>$ versus
the compact lattice momentum $<\vec{P}>$,
one is confronted with the problem of non-commuting operators.
Thus we have chosen to diagonalize the Hamiltonian,
considering a particular
energy $E$ and the corresponding eigen state $\psi_{E}$
and to compute $<\vec{P}>$ in this eigenstate.
As discussed in Ref. \cite{kn:Chaa92}, this eigenstate
is characterized by the quantum
number $k_{tot}$ the so-called "total wave vector"(see eq.(3)).
For small momenta one can show
$<\vec{P}> = C \vec{k}_{tot} + O(k^{2}_{tot})$,
where $C$ is a state-dependent factor.
Numerical results on this are given in Ref. \cite{kn:Chaa92}.
The non-linear corrections are expectred to drop off with the cut-off going to
infinity, i.e. $a \rightarrow 0$.
It is well known that on a finite lattice Lorentz symmetry is violated.
It should be restored in the continuum limit.
Restoration of Lorentz or Poincar\'{e} symmetry would imply
that $< M^{2} >$ is invariant. It would have been interesting
to compute the mass from $<M^{2}> = < H^{2} - \vec{P}^{2} >$.
However, a meaningful evaluation of $<\vec{P}^{2}>$ would have required
to work in Hilbert space with a number of plaquettes higher than four.

The results of the calculation of the dispersion relation
for the anti-symmetric glueball are presented in Fig.[1].
The wave functions involved in such a diagonalisation
are defined in the Hilbert space described above. The diagonalization was
performed numerically using IMSL.
The numerical data approximate Lorentz behavior reasonably well,
as is shown by a fit with a parabola.
Qualitatively similar results are obtained also for the
dispersion relation of the meson. The most serious source of
systematic error for such an object is the finite (and small)
size of the lattice. The magnitude of such an error can be
assessed by interpolating the continuous curve in Fig. [1]
to zero momentum and
by comparing with the value of the mass spectrum computed from
the diagonalization of the Hamiltonian using a zero-wave vector
basis. The data presented here have been extended from those of Ref.
\cite{kn:Chaa92}
to larger values of $x$ by using a larger number of plaquettes (up to
four) in the basis.

The results of the mass spectrum calculations are shown in Figs.[2-6].
Fig.[2] shows the ground state energy density
$ { \omega _0}/{N^2}$ as a function of the
coupling parameter $x$. Using a basis containing up to four plaquettes
is most accurate in the strong coupling region.
As a check, these results are compared with the calculation of Irving et al.
\cite{kn:Irvi83} who have used
Hamiltonian diagonalisation on a space lattice. Our data
are also compared with the strong coupling perturbation study of
Alessandrini et al. \cite{kn:Ales80}.

We now present our results for the rest mass. These were
calculated in the zero-wave vector sector,
subtracting out the vacuum energy.
The data of Fig.[3] correspond to the anti-symmetric glueball
on $5^2$ lattice.
This particular state has been studied
as a function of lattice size up to $11^{2}$. We find
a reasonably converged behavior starting at about a lattice size of $5^{2}$.
A similar behavior is found for the mass of the symmetric glueball.
The mass ratio of symmetric to anti-symmetric glueball is given in
Fig.[4]. Note that in the strong coupling
limit $g = \infty$, i.e. $x =0$, the symmetric and antisymmetric states are
degenerate.
Finally, in Fig.[5], we display the meson mass as a function of the coupling
parameter. This corresponds to an anti-symmetric meson state.
One observes a quite similar behavior as for the glueball.
Although our lattice calculation is
non-perturbative, it is based on a sector of Hilbert space. For smaller
values of the coupling constant $g$, i.e., larger coupling parameters $x$,
one needs to take into account higher excitations, e.g., electric flux
between the quarks of several lattice units and more than four plaquettes.
Comparing the mass spectrum extracted
from the interpolation of the dispersion relation
and from the direct diagonalization of the Hamiltonian  in the
zero-wave vector sector, we find a typical discrepancy of
two percent (Figs.[2],[5]), which is an estimate of finite size effects.
Those effects are thus under control in the range of the coupling parameter
considered in this work.\\

\paragraph{5. Scattering on the lattice.}
Key elements in the computation of scattering amplitudes
in a Hilbert space formulation are the generator of the time evolution,
i.e., the lattice Hamiltonian, and the initial and final states.
The latter are characterized by quantum numbers like energy, momentum,
spin etc. However, to describe in Hilbert space a reaction like
glueball+glueball $\rightarrow$ glueball+glueball,
quantum numbers alone are not sufficient; in addition one needs
to construct asymptotic one-particle states.
The way to do this has been proposed a long time ago
in papers by Haag and Ruelle \cite{kn:Haag58}.
Haag-Ruelle's theory specifies which operators give asymptotic
one-particle states when applied on the physical vacuum.
However, it does not answer how to obtain the physical vacuum state.
In the truncated Hilbert space, built from link and plaquette states,
we can compute the whole truncated spectrum.
Thus in particular, we obtain the ground state, as discussed in sect. 4,
and the next lowest eigenstate, which is a one-particle glueball state.
Thus in the truncated scheme, we are able to construct a Haag-Ruelle operator,
i.e., an explicit representation of an operator
which maps the vacuum state onto the glueball state.
An asymptotic two-particle state is obtained by applying a Haag-Ruelle operator
twice on the physical vacuum.
There are some tests on the construction of asymptotic one- or two-particle
states. Firstly, it is quite simple to verify that an asymptotic
one-particle state is an eigenstate of the Hamiltonian, but an asymptotic
two-particle state is not. We have verified this for glueballs and mesons.
Secondly, one can verify localisation.
I.e., if $\phi_{as}$ is an asymptotic one-particle state, then
\begin{equation}
<\phi_{as} \mid C(x) \mid \phi_{as}> = <0 \mid C(x) \mid 0> + \epsilon (x)
\end{equation}
for a local observable $C(x)$, where $\epsilon (x)$ tends to zero for
space-like distances $x \rightarrow 0$.
This we have not verified in this work, because
we have worked on relatively small lattices.
We defer it to future studies.
In this work an asymptotic one-particle state is given by
the glueball/ meson
state as discussed in sect.4, being the lowest lying states
above the vacuum in the glueball/meson
sector. Asymptotic two-particle states are product states of
the one-particle states.

A time-dependent Hilbert space method to compute $S$-matrix elements has been
described and reviewed in Ref. \cite{kn:Krog92}.
An $S$-matrix element is given by
\begin{equation}
S_{fi} =  \lim_{t \rightarrow \infty}
< \phi^{fi}_{as} \mid \exp [i E_{as} t] \exp[ -i 2 H t] \exp [i E_{as} t]
\mid \phi^{in}_{as} >,
\end{equation}
where $H$ denotes the Hamiltonian and $\phi^{fi}_{as}$, $\phi^{in}_{as}$
denote the asymptotic states.
On the lattice this is replaced by
\begin{equation}
S_{fi}(T) =
< \phi^{fi}_{as} \mid \exp [i E_{as} T] \exp[ -i 2 H(N) T] \exp [i E_{as} T]
\mid \phi^{in}_{as} >,
\end{equation}
where $H(N)$ denotes the finite dimensional lattice Hamiltonian
defined in the truncated Hilbert space.
The time-evolution $\exp [-i 2 H(N) T]$
is computed in the eigen representation of $H(N)$.
The reason to introduce as parameter a large but finite scattering time $T$
is based on the property that eq.(11) does not have a limit when
$T \rightarrow \infty$.
However, study of this and other models shows \cite{kn:Krog92}
that there is a range of the scattering parameter $T$, such that $S_{fi}(T)$ is
stable and gives physically meaningful results.
The value of $T$ is determined by the physical requirement
that energy violation of the scattering process measured on the lattice
by
\begin{equation}
\Delta_{<E>}(T) = \mid <\psi_{scatt}(T) \mid H(N) \mid \psi_{scatt}(T) >
-E_{as} \mid
\end{equation}
becomes minimal.
The property that the position of the minimum
of energy violation coincides with a region of stability of
$S_{fi}(T)$ can be seen in Fig.[6] for anti-symmetric glueball scattering.
A similar behavior is obtained for meson scattering.

By extracting the invariant matrix element and putting the suitable kinematical
factor, we have computed the differential cross sections for scattering of two
glueballs. The results are shown in Fig.[7].
Our results are compared with the predictions by Hakim \cite{kn:Haki81}
obtained from strong coupling perturbation theory and agree quite well.
Finally, in Fig.[8] we give the cross section for meson-meson scattering
which shows a quite similar behavior.
\\

In conclusion,
we have studied glueballs and mesons in compact $QED_{2+1}$ gauge theory
in a Hamiltonian formulation on a momentum latttice. Mesons are treated as
strings of electric field between classical quark-antiquark sources.
We have computed masses and find agreement with Hamiltonian calculations
on a coordinate lattice and with strong coupling perturbation theory.
The new aspects of this work are:
(a) In analogy to the Kogut-Susskind compact Hamiltonian,
we suggest a compact lattice momentum operator and compute glueball
and meson dispersion relations.
(b) Using a non-perturbative time-dependent Hilbert space method
we compute glueball and meson scattering cross sections. For glueballs
we compare our results in the strong coupling regime with perturbation theory
and find good agreement.

\newpage
\begin{flushleft}
{\bf Figure Captions}
\end{flushleft}
\begin{description}
\item[{Fig.1}]
Dispersion relation $E$ versus $<P_{x}>$
for anti-symmetric glueball state.
Coupling parameter: $x=0.26$. Lattice size: $5^{2}$. The curve is a fit
to the numerical data in order to extract the mass.
\item[{Fig.2}]
Ground state energy density versus coupling parameter $x$.
Full line: diagonalisation on a coordinate lattice by Irving et al.
\cite{kn:Irvi83}. Circle: strong coupling perturbation theory
by Alessandrini et al. \cite{kn:Ales80}. Triangle: this work. Lattice size:
$5^{2}$.
\item[{Fig.3}]
Mass of anti-symmetric glueball state. Otherwise same as Fig.[2].
\item[{Fig.4}]
Mass ratio of symmetric to anti-symmetric
glueball state. Otherwise same as Fig.[2].
\item[{Fig.5}]
Meson mass versus coupling parameter $x$.
Lattice size: $5^{2}$.
\item[{Fig.6}]
Scattering of anti-symmetric glueballs as function of scattering time parameter
$T$. Full line: imaginary part of S-matrix. Dashed line:
violation of energy conservation. Coupling parameter: $x=0.4$.
Lattice size: $5^{2}$.
\item[{Fig.7}]
Scattering of anti-symmetric glueballs.
Total croass section versus total momentum.
Full line: strong coupling perturbation theory by Hakim \cite{kn:Haki81}.
Triangle: this work. Coupling parameter: $x=0.4$. Lattice size: $5^{2}$.
\item[{Fig.8}]
Same as Fig.[7] for meson-meson scattering.
\end{description}
\end{document}